\documentclass[reprint,prc,showpacs,amsmath,amssymb,superscriptaddress]{revtex4-1}
\usepackage{dcolumn}
\usepackage[dvips]{graphicx}
\usepackage{amsmath}
\usepackage{multirow}
\usepackage{setspace}
\newcommand\T{\rule{0pt}{3.1ex}}
\usepackage{color}
\usepackage{bm}
\usepackage[T1]{fontenc}
\usepackage[latin9]{inputenc}
\usepackage{color}
\usepackage{amssymb}
\usepackage{float}

\providecommand{\tabularnewline}{\\}

\usepackage{array}

\begin{document}

\title{$0\nu\beta\beta$ and $2\nu\beta\beta$ nuclear matrix elements in the interacting boson model with isospin restoration}

\author{J.\ Barea}
\email{jbarea@udec.cl}
\affiliation{Departamento de F\'{i}sica, Universidad de Concepci\'{o}n,
 Casilla 160-C, Concepci\'{o}n 4070386, Chile}
 
\author{J. Kotila}
\email{jenni.kotila@yale.edu}
\affiliation{Center for Theoretical Physics, Sloane Physics Laboratory,
Yale University, New Haven, Connecticut 06520-8120, USA}
\affiliation{University of Jyvaskyla, Department of Physics, B.O. Box 35, FI-40014, University of Jyvaskyla, Finland}

\author{F.\ Iachello}
\email{francesco.iachello@yale.edu}
\affiliation{Center for Theoretical Physics, Sloane Physics Laboratory,
 Yale University, New Haven, Connecticut 06520-8120, USA}

\begin{abstract}
We introduce a method for isospin restoration in the calculation of nuclear matrix elements (NME) for $0\nu\beta\beta$ and $2\nu\beta\beta$ decay within the framework of interacting boson model (IBM-2). With this method, we calculate NME for all processes of interest in $0\nu\beta^-\beta^-$, $2\nu\beta^-\beta^-$, and in $0\nu\beta^+\beta^+$, $0\nu\beta^+ EC^+$, $R0\nu ECEC$, $2\nu\beta^+\beta^+$, $2\nu\beta^+EC$, and $2\nu ECEC$. With this method, the Fermi (F) matrix elements for $2\nu\beta\beta$ vanish, and those for $0\nu\beta\beta$ are considerably reduced. 
\end{abstract}

\maketitle

\section{Introduction}
\begin{figure}[h]
\includegraphics[width=7.6cm]{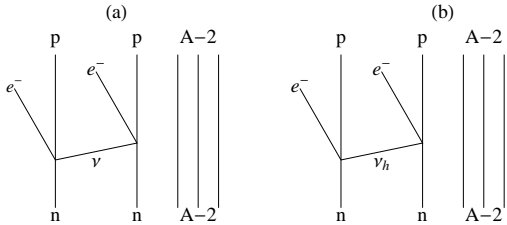} 
\caption{\label{fig1} Neutrinoless double-$\beta$ decay mechanism for (a) light neutrino exchange and (b) heavy neutrino exchange.}
\end{figure}

The question whether neutrinos are Majorana or Dirac particles,  and of what are their masses and phases in the mixing matrix remains one of the most important in physics today. A direct measurement of the average mass can be obtained from the observation of the neutrinoless double-$\beta$ decay ($0\nu\beta\beta$)
\begin{equation}
^A_ZX^N \rightarrow ^A_{Z\pm2}Y_{N\mp2}+2e^{\mp},
\end{equation}
two scenarios of which  are shown in Fig.~\ref{fig1}
Several experiments are underway to detect this decay, and others are in the planning stage (for a review, see  e.g. \cite{cremonesi}). The half-life for this decay can be written as 
\begin{equation}
\label{hl0}
\lbrack\tau_{1/2}^{0\nu}]^{-1}=G_{0\nu}\left\vert M_{0\nu}\right\vert ^{2}\left\vert f(m_{i},U_{ei})\right\vert ^{2},
\end{equation}
where $G_{0\nu}$ is a phase space factor, $M_{0\nu}$ the nuclear
matrix element and $f(m_{i},U_{ei})$ contains physics beyond the
standard model through the masses $m_{i}$ and mixing matrix elements
$U_{ei}$ of neutrino species.

\begin{figure}[h]
\begin{center}
\includegraphics[width=3.5cm]{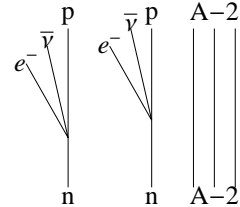} 
\end{center}
\caption{\label{fig2} Double-$\beta$ decay mechanism with the emission of $2\bar{\nu}$.}
\end{figure}

Concomitant with the neutrinoless modes, there is also the process allowed by the standard model, $2\nu\beta\beta$, depicted in Fig.~\ref{fig2}. For this process, the half-life can be, to a good approximation, factorized in the form
\begin{equation}
\label{hl}
\left[\tau_{1/2}^{2\nu}\right]^{-1}=G_{2\nu}\left\vert m_ec^2M_{2\nu}\right\vert ^{2}.
\end{equation}
The factorization here is not exact and conditions under which
it can be done are discussed in Ref.~\cite{kotila}.
 
The processes depicted in Figs.~\ref{fig1} and \ref{fig2} are of the type 
\begin{equation}
\left(A,Z\right)\rightarrow\left(A,Z+2\right)+2e^{-}+\text{anything.}
\end{equation}
 In  recent years, interest in the processes 
\begin{equation}
\left(A,Z\right)\rightarrow(A,Z-2)+2e^{+}+\text{anything}
\end{equation}
 has also arisen. In this case there are also the competing modes
in which either one or two electrons are captured from the electron
cloud (0$\nu\beta$EC, 2$\nu\beta$EC, $R0\nu$ECEC, 2$\nu$ECEC). Also for these
modes, the half-life can be factorized (either exactly or approximately)
into the product of a phase space factor and a nuclear matrix element
which then are the crucial ingredients of any double-$\beta$ decay calculation.

In order to extract physics beyond the standard model, contained in the function $f$ in Eq.~(\ref{hl0}), we need an accurate calculation of both $G_{0\nu}$ and $M_{0\nu}$. These calculations will serve the purpose of extracting the neutrino mass $\langle m_{\nu}\rangle$ if $0\nu\beta\beta$ is observed, and of guiding searches if $0\nu\beta\beta$ is not observed.

Recently we have started a systematic evaluation of phase space
factors (PSF) and nuclear matrix elements (NME)
 for all processes of interest. The results for NME have been presented in Refs.~\cite{barea,barea12,barea13,barea13b,kotila14}, and those for PSF in Refs. \cite{kotila,kotila14,kotila13}. The calculations for the nuclear matrix elements have been done within the framework of the microscopic interacting boson model (IBM-2).

Having completed the calculations in all nuclei of interest, we have now readdressed them with the purpose of providing as accurate as possible results. As shown in Table XV of Ref. \cite{barea13}, the Fermi matrix elements $M^{(2\nu)}_F$ for $2\nu\beta\beta$ decay in IBM-2 do not vanish in cases where protons and neutrons occupy the same major shell. Similarly, the Fermi matrix elements $M^{(0\nu)}_F$ for $0\nu\beta\beta$ decay are large when protons and neutrons are in the same major shell, as we can see from Table VII of Ref. \cite{barea13}, where the quantity $\chi_F=(g_V/g_A)^2 M_F^{(0\nu)}/M_{GT}^{(0\nu)}$ is reported. The same problem with isospin was present in the quasiparticle random phase approximation (QRPA) both for QRPA-T\"{u} \cite{x} and QRPA-Jy \cite{xx} and it was cured recently \cite{sim13} by changing the values of the renormalization constant $g_{pp}^{T=1}$ which is adjusted in such a way as to make $M^{(2\nu)}_F$ vanish. In this paper, we report on a method similar in spirit, but different in practice from QRPA, and use it to impose the condition $M^{(2\nu)}_F=0$ in IBM-2. A consequence of the implementation of this method is that the matrix elements $M^{(0\nu)}_F$ are reduced and comparable now to those obtained in the Interacting Shell Model (ISM).



\section{Formalism}
The role of isospin in the Interacting Boson Model was extensively investigated in the 1980's and 1990's \cite{ell80,ell81,ell87,ell87b, hal84} and is summarized in the paper "Isospin and F-spin in the Interacting Boson Model" by J. P Elliott \cite{ell90}. As discussed in \cite{ell90}, IBM-2 wave functions have good isospin in heavy nuclei with a neutron excess. The problem arises only in light nuclei where protons and neutrons occupy the same orbits ($sd$ and $pf$ shells). For these nuclei one needs to introduce an isospin invariant form of IBM, called IBM-3 \cite{ell80}. IBM-2 can still be used in light nuclei if the parameters of the interaction are obtained by projection from those of the isospin invariant IBM-3. (For the nuclei discussed in this paper, only $^{48}$Ca and $^{48}$Ti are in a situation in which IBM-3 or a projected form should be used.) As a result, isospin does not pose a problem for IBM-2 wave functions.

The problem with isospin in IBM-2 arises in the mapping of the fermion operator for $0\nu\beta\beta$ and $2\nu\beta\beta$ decay. The expression for the fermionic transition operator of type Fermi (F), Gamow-Teller (GT), and tensor (T) is \cite{barea}
\begin{equation}
\label{eq6}
V_{s_1, s_2}^{(\lambda)}=\frac{1}{2}\sum_{n,n'}\tau^+_n \tau^+_{n'}\left[\Sigma_n^{s_1}\times \Sigma_{n'}^{s_2}\right]^{(\lambda)}\cdot V(r_{nn'})C^{(\lambda)}(\Omega_{nn'}),
\end{equation}
where, for $s=0$, $\Sigma^{(0)}=1$, and for $s=1$, $\Sigma^{(1)}=\vec{\sigma}$. In particular, the Fermi transition operator for $2\nu\beta\beta$ decay, obtained from Eq. (\ref{eq6}) by setting $\lambda=0$, $s_1=s_2=0$, and $V(r)=1$, is
\begin{equation}
\label{eq7}
V^{(2\nu)}_F=\frac{1}{2}\sum_{n,n'}\tau^+_n \tau^+_{n'} \hspace{1cm} (2\nu\beta\beta)
\end{equation}
and for $0\nu\beta\beta$ decay obtained from Eq. (\ref{eq6}) by setting $\lambda=0$, $s_1=s_2=0$, and $V(r)=H(r)$, is 
\begin{equation}
V^{(0\nu)}_F=\frac{1}{2}\sum_{n,n'}\tau^+_n \tau^+_{n'}H(r_{nn'}) \hspace{1cm} (0\nu\beta\beta)
\end{equation}
where $H(r)$ is given in Appendix A of Ref. \cite{barea}. The operator (\ref{eq7}), when summed over all particles, cannot change isospin and its matrix elements between initial and final states must vanish. 

In previous IBM-2 calculations \cite{barea13}, the matrix elements $M_F^{(2\nu)}$ were simply discarded and the matrix elements $M_F^{(0\nu)}$ were kept untouched. We suggest here that a better approximation is that of modifying the mapped operator by imposing the condition that $M_F^{(2\nu)}=0$. This condition can be simply implemented in our calculation by replacing the radial integrals of Appendix A of Ref. \cite{barea},  $R^{(k_1,k_2,\lambda)}(n_1,l_1,n_2,l_2,n'_1,l'_1,n'_2,l'_2)$, by

\begin{equation}
\begin{split}
&2\nu\beta\beta: \\
&R^{(k_1,k_2,\lambda)}(n_1,l_1,n_2,l_2,n_1',l_1',n_2',l_2')\\
&-\delta_{k_1,0}\delta_{k_2,0}\delta_{k,0}\delta_{\lambda,0}\delta_{j_1,j_1'}\delta_{j_2,j_2'}\delta_{n_1,n_1'}\delta_{l_1,l_1'}\delta_{n_2,n_2'}\delta_{l_2,l_2'}\\
\end{split}
\end{equation}

\begin{equation}
\begin{split}
&0\nu\beta\beta: \\
&R^{(k_1,k_2,\lambda)}(n_1,l_1,n_2,l_2,n_1',l_1',n_2',l_2')\\
&-\delta_{k_1,0}\delta_{k_2,0}\delta_{k,0}\delta_{\lambda,0}\delta_{j_1,j_1'}\delta_{j_2,j_2'}\delta_{n_1,n_1'}\delta_{l_1,l_1'}\delta_{n_2,n_2'}\delta_{l_2,l_2'}\\
&\times R_{0\nu}^{(0,0,0)}(n_1,l_1,n_2,l_2,n_1',l_1',n_2',l_2')
\end{split}
\end{equation}
where 
\begin{equation}
\begin{split}
R_{0\nu}^{(0,0,0)}&(n_1,l_1,n_2,l_2,n_1',l_1',n_2',l_2')\\
=&\int_0^\infty \frac{2}{\pi}\frac{1}{p(p+\tilde{A})}p^2dp \\
\times&\int_0^\infty R_{n_1l_1}(r_1)R_{n_1'l_1'}(r_1)\frac{\sin(pr_1)}{p_1^2}r_1^2dr_1\\
\times\ &\int_0^\infty R_{n_2l_2}(r_2)R_{n_2'l_2'}(r_2)\frac{\sin(pr_2)}{p_2^2}r_2^2dr_2.\\
\end{split}
\end{equation}

This prescription will guarantee that the F matrix elements vanish for $2\nu\beta\beta$, and will reduce the F matrix elements for $0\nu\beta\beta$ by subtraction of $R_{0\nu}^{(0,0,0)}$, which is the monopole term in the expansion of the matrix element into multipoles. It is similar to the prescription used in Ref.~\cite{sim13} (see Fig. 4 of Ref.~\cite{sim13}).

Although the method described in this section is not an isospin restoration of the IBM-2 wave functions which have already good isospin but rather a restoration of the isospin properties of the mapping of the transition operator, we shall, nonetheless, for simplicity refer to it in the following sections as "simply isospin restoration".

\section{Results for $0\nu\beta\beta$}

From here on, the calculation of the matrix elements in the interacting boson model proceeds in the same way as in Refs. \cite{barea,barea13}. We do not repeat the formulas but only report the results of the calculation. In the calculation one needs to specify the parametrization of short-range correlations (SRC). In earlier calculations \cite{barea} the Miller-Spencer parametrization was used. It has now become clear that Argonne and CD-Bonn parametrizations are more appropriate. Here we use throughout the Argonne parametrization of the correlation function
\begin{equation}
f(r)=1-ce^{ar^2}(1-br^2),
\end{equation}
that is $a=1.59$fm$^{-2}$, $b=1.45$fm$^{-2}$, and $c=0.92$. We write
\begin{equation}
\begin{split}
\label{sum}
M_{0\nu} & =  g_{A}^{2}M^{(0\nu)}, \\
M^{(0\nu)} & =  M_{GT}^{(0\nu)} -\left(\frac{g_{V}}{g_{A}}\right)^2 M_{F}^{(0\nu)}+M_{T}^{(0\nu)},
\end{split}
\end{equation}
with the ratio $g_V/g_A$ explicitly displayed in front of $M_F^{(0\nu)}$.

\subsection{$0\nu\beta\beta$ decay with light neutrino exchange}

In Table~\ref{table1}, we show the results of our calculation of the nuclear matrix elements to the ground state, $0^+_1$, and the first excited state,  $0^+_2$, broken down into GT, F, and T contributions and their sum according to Eq. (\ref{sum}). The parameters of the IBM-2 Hamiltonian used in this calculation are those given in Table XXIII of Ref.~\cite{barea13} (with the exception of $^{154}$Gd whose parameters are given in Ref.~\cite{bel13}).

\begin{table*}[cbt!]
 \caption{\label{table1}IBM-2 nuclear matrix elements $M^{(0\nu)}$ (dimensionless) for $0\nu\beta^-\beta^-$ decay with Argonne SRC, $g_V/g_A=1/1.269$, and isospin restoration.}
 \begin{ruledtabular} %
\begin{tabular}{ccccccccc}
&\multicolumn{4}{c}{$0_{1}^{+}$} &\multicolumn{4}{c}{$0_{2}^{+}$}\\\cline{2-5}\cline{6-9}
\T
$A$  & $M_{GT}^{(0\nu)}$  & $M_{F}^{(0\nu)}$  & $M_{T}^{(0\nu)}$  & $M^{(0\nu)}$  & $M_{GT}^{(0\nu)}$  & $M_{F}^{(0\nu)}$  & $M_{T}^{(0\nu)}$  & $M^{(0\nu)}$\tabularnewline
\hline 
\T
$^{48}$Ca  & 1.73  & -0.30  & -0.17  & 1.75   		& 3.78  & -0.27  & -0.12  & 3.82\tabularnewline
$^{76}$Ge  & 4.49  & -0.68  & -0.23  & 4.68  		& 1.95  &-0.27   & -0.09  & 2.02\tabularnewline
$^{82}$Se  & 3.59  & -0.60  & -0.23  & 3.73 		& 0.92  &-0.13   & -0.05  & 0.95\tabularnewline
$^{96}$Zr  & 2.51  & -0.33  & 0.11  & 2.83  		& 0.04  &-0.01   & 0.00  & 0.05\tabularnewline
$^{100}$Mo  & 3.73  & -0.48  & 0.19  & 4.22  		& 0.99  &-0.13   & 0.05  & 1.12\tabularnewline
$^{110}$Pd  & 3.59  & -0.40  &0.21   & 4.05  		& 0.46  &-0.05   & 0.03  & 0.52\tabularnewline
$^{116}$Cd  & 2.76  & -0.33  &0.14   & 3.10 		& 0.84  &-0.09   & 0.03  & 0.93\tabularnewline
$^{124}$Sn  & 2.96  & -0.57  & -0.12  & 3.19  		& 2.21 & -0.41  & -0.09  & 2.38\tabularnewline
$^{128}$Te  & 3.80  & -0.72  & -0.15  & 4.10  		& 2.65 & -0.47  & -0.09  & 2.85\tabularnewline
$^{130}$Te  & 3.43  & -0.65  & -0.13  & 3.70  		& 2.52 & -0.45  & -0.08  & 2.71\tabularnewline
$^{134}$Xe   &3.77 &-0.68  &-0.15  &4.05 	      	&2.19  &-0.36   &-0.06   &2.35\tabularnewline
$^{136}$Xe  & 2.83  & -0.52  & -0.10  & 3.05  		& 1.49 &-0.24   & -0.03  & 1.60\tabularnewline
$^{148}$Nd  & 2.00  & -0.38  & 0.07  & 2.31  		& 0.25  & -0.05  & 0.01  & 0.29\tabularnewline
$^{150}$Nd  & 2.33  & -0.39  & 0.10  & 2.67  		& 0.40  &-0.06   & 0.02  & 0.45\tabularnewline
$^{154}$Sm  & 2.49  & -0.36  & 0.11  & 2.82  		& 0.37  &-0.04   & 0.01  & 0.41\tabularnewline
$^{160}$Gd  & 3.64  & -0.45  & 0.17  & 4.08  		& 0.76  &-0.11   & 0.04  & 0.87\tabularnewline
$^{198}$Pt  & 1.90 & -0.33  & 0.09 & 2.19  		& 0.08  &-0.02   & 0.01  & 0.10\tabularnewline
$^{232}$Th  & 3.58  & -0.44  & 0.18  & 4.04  		& 0.12  &-0.02   & 0.01  & 0.15\tabularnewline
$^{238}$U  & 4.27  & -0.53  & 0.21  & 4.81  		& 0.34  &-0.05   & 0.02  & 0.40\tabularnewline
\end{tabular}\end{ruledtabular} 
\end{table*}

\begin{figure}[h]
\begin{center}
\includegraphics[width=8.6cm]{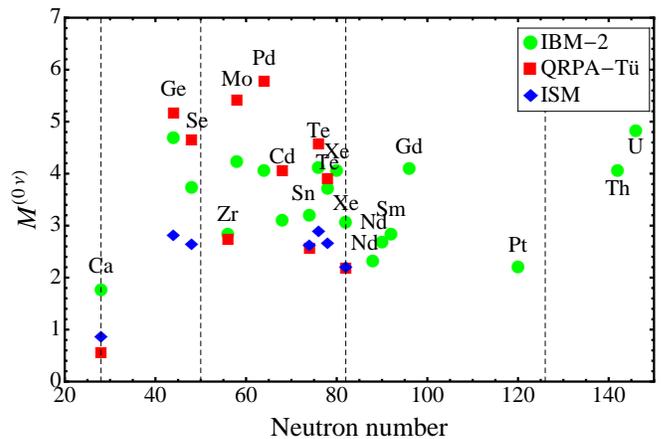} 
\end{center}
\caption{\label{fig3}(Color online) IBM-2 isospin restored results for $0\nu\beta^-\beta^-$ decay compared with QRPA-T\"{u} \cite{sim13} and the ISM \cite{poves}.}
\end{figure}

When compared with the matrix elements without the restoration \cite{barea13}, we see a considerable reduction of the F matrix elements to values comparable to those of the shell model. The overall reduction in $M^{(0\nu)}$ is $\sim 15\%$. Our results are compared with QRPA-T\"{u} with isospin restoration (Argonne SRC) \cite{sim13} and ISM (UCOM SRC) \cite{poves} in Table~\ref{table2} and Fig.~\ref{fig3}.
\begin{ruledtabular}
\begin{center}
\begin{table}[cbt!]
\caption{\label{table2}
Comparison among nuclear matrix elements for $0\nu\beta^-\beta^-$ decay to ground state, $0_{1}^{+}$, in IBM-2 with Argonne SRC, $g_A=1.269$, and isospin restoration,  QRPA-T\"{u} with Argonne SRC, $g_A=1.27$, and isospin restoration \cite{sim13}, and ISM with UCOM SRC and $g_A=1.25$ \cite{poves}. All matrix elements are in dimensionless units.}
\begin{tabular}{cccc}
   &   &  \ensuremath{M^{(0\nu)}} &  \\\cline{2-4}
   \T
Decay &  IBM-2  &  QRPA-T\"u  &  ISM \\
\hline
\T
$^{48}$Ca$\rightarrow ^{48}$Ti		&1.75	&0.54	 &0.85\\
$^{76}$Ge$\rightarrow ^{76}$Se	&4.68	&5.16 	&2.81\\
$^{82}$Se$\rightarrow ^{82}$Kr		&3.73	&4.64 	&2.64\\
$^{96}$Zr$\rightarrow ^{96}$Mo	&2.83	&2.72 	&\\
$^{100}$Mo$\rightarrow ^{100}$Ru 	&4.22	&5.40 	&\\
$^{110}$Pd$\rightarrow ^{110}$Cd 	&4.05	&5.76	 &\\
$^{116}$Cd$\rightarrow ^{116}$Sn 	&3.10	&4.04 	&\\
$^{124}$Sn$\rightarrow ^{124}$Te 	&3.19	&2.56	 &2.62\\
$^{128}$Te$\rightarrow ^{128}$Xe 	&4.10	&4.56	&2.88\\
$^{130}$Te$\rightarrow ^{130}$Xe 	&3.70	&3.89	&2.65\\
$^{134}$Xe$\rightarrow ^{134}$Ba 	&4.05	&		&\\
$^{136}$Xe$\rightarrow ^{136}$Ba 	&3.05	&2.18	&2.19\\
$^{148}$Nd$\rightarrow ^{148}$Sm 	&2.31	&		&\\
$^{150}$Nd$\rightarrow ^{150}$Sm 	&2.67	&		&\\
$^{154}$Sm$\rightarrow ^{154}$Gd 	&2.82	&		&\\
$^{160}$Gd$\rightarrow ^{160}$Dy 	&4.08	&		&\\
$^{198}$Pt$\rightarrow ^{198}$Hg 	&2.19	&		&\\
$^{232}$Th$\rightarrow ^{232}$U 	&4.04 	& 		&\\
$^{238}$U$\rightarrow ^{238}$Pu 	&4.81 	&		&\\
\end{tabular}
\end{table}
\end{center}
\end{ruledtabular}

The reduction in the Fermi matrix elements $M_F^{(0\nu)}$ brought in by the isospin restoration is shown in Table~\ref{chitable} where the quantity $\chi_F=(g_V/g_A)^2 M^{(0\nu)}_F/M^{(0\nu)}_{GT}$ is shown for the old and new calculation and compared  with QRPA without (old )and with (new) isospin restoration, and with  ISM.  Our isospin restored Fermi matrix  elements are comparable to those of the ISM, but a factor of 2-3 smaller than the isospin restored QRPA-T\"{u} results. This may be due to the fact that both in IBM-2 and ISM the model space is rather restricted, while in QRPA several major shells are included.
\begin{ruledtabular}
\begin{center}
\begin{table}[h]
\caption{\label{chitable}Comparison between Fermi matrix elements, $\chi_F$, for $0\nu\beta^-\beta^-$ decay in IBM-2, QRPA-T\"{u} \cite{sim13} and ISM \cite{poves,caurier2007}.}
\begin{tabular}{cccccc}
&\multicolumn{5}{c}{$\chi_F$} \\\cline{2-6}
\T
Decay      		&\multicolumn{2}{c}{IBM-2}  &\multicolumn{2}{c}{QRPA-T\"{u}}   &  ISM \\
&old &new &old &new \\\cline{2-3} \cline{4-5}
\T
$^{48}$Ca		&-0.39	&-0.11	&-0.93		&-0.32		&-0.14	\\
$^{76}$Ge	&-0.37	&-0.09	&-0.34		&-0.21		&-0.10	\\
$^{82}$Se		&-0.40	&-0.10	&-0.35		&-0.23		&-0.10	\\
$^{96}$Zr		&-0.08	&-0.08	&-0.38		&-0.23		&	\\
$^{100}$Mo	&-0.08	&-0.08	&-0.30		&-0.30		&	\\
$^{110}$Pd	&-0.07	&-0.07	&-0.33		&-0.27		&-0.16	\\
$^{116}$Cd	&-0.07	&-0.07	&-0.30		&-0.30		&-0.19	\\
$^{124}$Sn	&-0.34	&-0.12	&-0.40		&-0.27		&-0.13	\\
$^{128}$Te	&-0.33	&-0.12	&-0.38		&-0.27		&-0.13	\\
$^{130}$Te	&-0.33	&-0.12	&-0.39		&-0.27		&-0.13	\\
$^{134}$Xe	&		&-0.11	&		&			&	\\
$^{136}$Xe	&-0.11	&-0.11	&-0.38		&-0.25		&-0.13	\\
$^{148}$Nd	&-0.12	&-0.12	&		&			&	\\
$^{150}$Nd	&-0.10	&-0.10	&		&			&	\\
$^{154}$Sm	&-0.09	&-0.09	&		&			&	\\
$^{160}$Gd	&-0.08	&-0.08	&		&			&	\\
$^{198}$Pt	&-0.11	&-0.11	&		&			&	\\
$^{232}$Th	&-0.08	&-0.08	&		&			&\\
$^{238}$U		&-0.08	&-0.08	&		&			&\\
\end{tabular}
\end{table}
\end{center}
\end{ruledtabular}

\subsection{$0\nu\beta\beta$ decay with heavy neutrino exchange}
These matrix elements can be simply calculated by replacing the potential
$v(p)=2\pi^{-1}[p(p+\tilde{A})]^{-1}$ in $R^{(k_1,k_2,\lambda)}$
by the potential $v_{h}(p)=2\pi^{-1}(m_{e}m_{p})^{-1}$. Table~\ref{table4} gives the corresponding matrix
elements. The index \textquotedbl{}h\textquotedbl{} distinguishes
these matrix elements from those with light neutrino exchange.
\begin{table*}[cbt!]
 \caption{\label{table4}Nuclear matrix elements for the heavy neutrino exchange mode of the
neutrinoless double-$\beta^-$ decay to the ground state (columns 2,3,4,
and 5) and to the first excited state (columns 6,7,8, and 9) using
the microscopic interacting boson model (IBM-2) with isospin restoration and Argonne SRC and $g_V/g_A=1/1.269$.}
 \begin{ruledtabular} %
\begin{tabular}{ccccccccc}
&\multicolumn{4}{c}{$0_{1}^{+}$} &\multicolumn{4}{c}{$0_{2}^{+}$}\\ \cline{2-5} \cline{6-9}
\T
$A$  & $M_{GT}^{(0\nu_h)}$  & $M_{F}^{(0\nu_h)}$  & $M_{T}^{(0\nu_h)}$  & $M^{(0\nu_h)}$  & $M_{GT}^{(0\nu_h)}$  & $M_{F}^{(0\nu_h)}$  & $M_{T}^{(0\nu_h)}$  & $M^{(0\nu_h)}$\tabularnewline
\hline 
\T
$^{48}$Ca  & 53.5  &-23.2  	&-21.3  	& 46.6  		&44.8&-8.8&-6.5&43.7\tabularnewline
$^{76}$Ge  & 104  &-42.8  	&-26.9  	& 104  		&38.6&-14.9&-9.8&38.1\tabularnewline
$^{82}$Se  & 87.2  &-37.1   	&-27.3  	& 82.9  		&16.8&-6.5&-4.6&16.2\tabularnewline
$^{96}$Zr  & 67.9   &-29.2  	& 12.7  	& 98.7  		&1.4&-0.6&0.3&2.1\tabularnewline
$^{100}$Mo  & 111   &-46.8  	& 24.2  	& 164  		&29.3&-12.4&6.4&43.3\tabularnewline
$^{110}$Pd  & 100  &-41.4   	& 27.7  	& 154  		&13.5&-5.6&3.8&20.9\tabularnewline
$^{116}$Cd  & 73.9  &-31.2   	& 16.9  	& 110  		&18.0&-7.5&3.5&26.1\tabularnewline
$^{124}$Sn  & 73.7  &-33.1   	&-14.9  	& 79.3  		&50.1&-22.2&-9.9&54.0\tabularnewline
$^{128}$Te  & 93.4  &-41.7   	&-18.3  	& 101  		&55.7&-24.4&-10.3&60.5\tabularnewline
$^{130}$Te  & 84.8  &-37.9   	&-16.6  	& 91.8  		&51.5 &-22.6&-9.3&56.2\tabularnewline
$^{134}$Xe   &86.6  &-39.3      &-19.8      &91.2        		&38.7  &-17.3   &-7.9     &41.5\tabularnewline
$^{136}$Xe  & 66.8  &-29.7  	&-12.7  	& 72.6  		&25.6&-11.0&-4.1&28.3\tabularnewline
$^{148}$Nd  & 72.8  &-32.7   	& 9.6  	& 103  		&8.1&-3.7&1.0&11.4\tabularnewline
$^{150}$Nd  & 81.1  &-35.6   	& 13.2  	& 116  		&12.2&-5.3&1.8&17.3\tabularnewline
$^{154}$Sm  & 78.1  &-33.7   	& 13.8  	& 113  		&8.9&-3.8&1.2&12.4\tabularnewline
$^{160}$Gd  & 106  &-44.6  	& 21.5  	& 155  		&26.7&-11.4&6.2&40.0\tabularnewline
$^{198}$Pt  & 71.4  &-31.9   	& 12.8  	& 104  		&4.0&-1.8&0.9&6.1\tabularnewline
$^{232}$Th  & 107  &-44.0   	& 24.4  	& 159  		&6.2&-2.7&1.9&9.9\tabularnewline
$^{238}$U  & 127  &-52.5  	& 28.7  	& 189  		&12.7&-5.4&3.4&19.4\tabularnewline
\end{tabular}\end{ruledtabular} 
\end{table*}
Our results are compared with results of QRPA-T\"u \cite{fae14} and ISM \cite{nea14} in Table~\ref{table5}. 
\begin{ruledtabular}
\begin{center}
\begin{table}[cbt!]
\caption{\label{table5}Neutrinoless double-$\beta^-$ decay nuclear matrix elements to ground state, $0_{1}^{+}$, in IBM-2 with isospin restoration, Argonne SRC and $g_V/g_A=1/1.269$, QRPA-T\"u with isospin restoration and Argonne SRC \cite{fae14}, and ISM with UCOM SRC \cite{nea14} for the heavy neutrino exchange mode. All matrix elements are in dimensionless units.}
\begin{tabular}{cccc}
&\multicolumn{2}{c}{  \ensuremath{M_{h}^{(0\nu)}}}   \\ \cline{2-4}
\T
Decay  &  IBM-2  &  QRPA-T\"u   &ISM\\
 \hline
 \T
$^{48}$Ca$\rightarrow ^{48}$Ti		&46.6	&40	 		&47.5\\
$^{76}$Ge$\rightarrow ^{76}$Se	&104		&287		&138 	\\
$^{82}$Se$\rightarrow ^{82}$Kr		&82.9	&262 		&127	\\
$^{96}$Zr$\rightarrow ^{96}$Mo	&98.7	&184	 	&	\\
$^{100}$Mo$\rightarrow ^{100}$Ru 	&164		&342 		&\\
$^{110}$Pd$\rightarrow ^{110}$Cd 	&154		&333	 		&\\
$^{116}$Cd$\rightarrow ^{116}$Sn 	&110		&209	 		&\\
$^{124}$Sn$\rightarrow ^{124}$Te 	&79.3	&184	 		&\\
$^{128}$Te$\rightarrow ^{128}$Xe 	&101		&302			&\\
$^{130}$Te$\rightarrow ^{130}$Xe 	&91.8	&264			&\\
$^{134}$Xe$\rightarrow ^{134}$Ba 	&91.2	&			&\\
$^{136}$Xe$\rightarrow ^{136}$Ba 	&72.6	&152			&\\
$^{148}$Nd$\rightarrow ^{148}$Sm 	&103		&			&\\
$^{150}$Nd$\rightarrow ^{150}$Sm 	&116		&			&\\
$^{154}$Sm$\rightarrow ^{154}$Gd 	&113		&			&\\
$^{160}$Gd$\rightarrow ^{160}$Dy 	&155		&			&\\
$^{198}$Pt$\rightarrow ^{198}$Hg 	&104		&			&\\
$^{232}$Th$\rightarrow ^{232}$U 	&159		& 		&	\\
$^{238}$U$\rightarrow ^{238}$Pu 	&189  	&		&	\\
\end{tabular}
\end{table}
\end{center}
\end{ruledtabular}

\subsubsection{Sensitivity to parameter changes, model assumptions and operator
assumptions}

The sensitivity of IBM-2 to parameter changes, model assumptions, and operator assumptions was discussed in great detail in Ref.~\cite{barea13}. We do not repeat this discussion, but only note that because of isospin restoration the sensitivity of F matrix elements to isospin purity decreases from $10\%$ to $1\%$, including the case of $^{48}$Ca decay, which previously was treated separately from the others. Our error estimate for $0\nu\beta\beta$ is now $16\%$ for all nuclei. In the case of $0\nu_h\beta\beta$ we also estimate a reduced sensitivity of F matrix elements from $10\%$ to $1\%$, and a reduced sensitivity of the matrix elements F + GT to SRC from $50\%$ to $25\%$. This sensitivity is estimated by comparing matrix elements with Argonne-CD-Bonn and UCOM SRC. Our total error estimate for $0\nu_h\beta\beta$ is now $28\%$ for all nuclei. Our final matrix elements, with error estimate are given in Table~\ref{final}.

 \begin{table}[h]
 \caption{\label{final}Final double-$\beta^-$ decay IBM-2 matrix elements with isospin restoration, Argonne SRC and error estimate.}
 \begin{ruledtabular} %
\begin{tabular}{ccc}
Decay  & Light neutrino exchange & Heavy neutrino exchange \tabularnewline
\hline 
\T
$^{48}$Ca  &1.75(28)		&47(13)		  \tabularnewline
$^{76}$Ge  &4.68(75)		&104(29)		   	 \tabularnewline
$^{82}$Se  &3.73(60)		&83(23)		   	 \tabularnewline
$^{96}$Zr	&2.83(45)		        &99(28)		   	 \tabularnewline
$^{100}$Mo &4.22(68)		&164(46)	   	  \tabularnewline
$^{110}$Pd &4.05(65)		&154(43)		   	 \tabularnewline
$^{116}$Cd &3.10(50)		&110(31)		   	 \tabularnewline
$^{124}$Sn &3.19(51)		&79(22)		   	 \tabularnewline
$^{128}$Te &4.10(66)		&101(28)		   	 \tabularnewline
$^{130}$Te &3.70(59)		&92(26)		   	 \tabularnewline
$^{134}$Xe &4.05(65)		&91(26)	   	  \tabularnewline
$^{136}$Xe &3.05(59)		&73(20)	   	  \tabularnewline
$^{148}$Nd &2.31(37)		&103(29)		   	 \tabularnewline
$^{150}$Nd &2.67(43)		&116(32)		   	 \tabularnewline
$^{154}$Sm &2.82(45)		&113(32)		   	 \tabularnewline
$^{160}$Gd &4.08(65)		&155(43)		   	 \tabularnewline
$^{198}$Pt &2.19(35)		&104(29)		   	 \tabularnewline
$^{232}$Th &4.04(65)		&159(45)		   	 \tabularnewline
$^{238}$U &4.81(77)			&189(53)		   	 \tabularnewline
\end{tabular}\end{ruledtabular} 
\end{table}

\begin{figure}[h]
\begin{center}
\includegraphics[width=8.6cm]{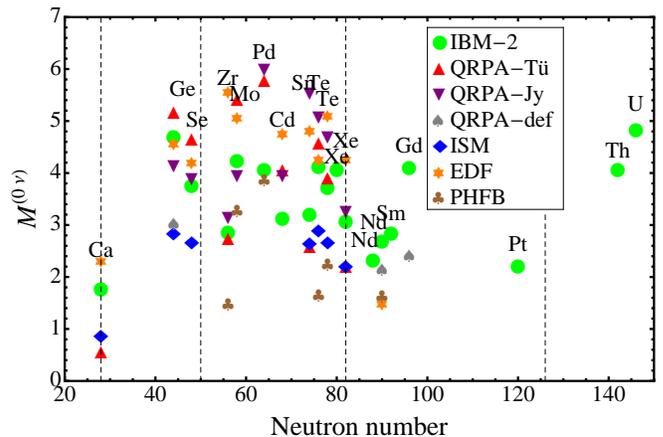} 
\end{center}
\caption{\label{allnme}(Color online) IBM-2 (Argonne) results for $0\nu\beta^-\beta^-$ nuclear matrix elements compared with QRPA-T\"u (Argonne) \cite{sim13}, ISM (UCOM) \cite{poves}, QRPA-Jy (UCOM) \cite{suh, suhpriv}, QRPA-deformed (CD-Bonn) \cite{fan11}, DFT  (UCOM) \cite{Martinez-Pinedo}, and HFB  (M-S) \cite{chandra}.}
\end{figure}

In addition to IBM-2, QRPA, and ISM, other calculations have been recently done. In Fig.~\ref{allnme} we compare our results with all available calculations including density functional theory (DFT) \cite{Martinez-Pinedo} and Hartree-Fock-Bogoliubov (HFB) theory \cite{chandra}.

\section{Results for $0\nu\beta^+\beta^+$}
Matrix elements for double positron ($\beta^+\beta^+$) decay and the related processes ($EC\beta^+$) and ($ECEC$) can be calculated in a similar way.

\subsection{$0\nu\beta^+\beta^+$ and related processes with light neutrino exchange}

In Table~\ref{0+} we show the results of our calculation of the matrix elements to the ground state, $0^+_1$, and first excited, $0^+_2$, state broken down into GT, F, and T contributions and their sum according to Eq. (\ref{sum}).

\begin{ruledtabular}
\begin{table*}[cbt!]
\caption{\label{0+}Nuclear matrix elements $M^{(0\nu)}$ (dimensionless) for neutrinoless $\beta^{+}\beta^{+}$, $EC\beta^{+}$, and $ECEC$ decays with Argonne SRC and $g_V/g_A=1/1.269$, in IBM-2 with isospin restoration.}
\begin{tabular}{lcccccccc}
 & \multicolumn{4}{c}{$0_{1}^{+}$} & \multicolumn{4}{c}{$0_{2}^{+}$}\tabularnewline \cline{2-5} \cline{6-9}
\T
Nucleus& $M_{GT}^{(0\nu)}$ & $M_{F}^{(0\nu)}$  & $M_{T}^{(0\nu)}$ & $M^{(0\nu)}$ & $M_{GT}^{(0\nu)}$ & $M_{F}^{(0\nu)}$  & $M_{T}^{(0\nu)}$ & $M^{(0\nu)}$\tabularnewline
\hline 
\T
$^{58}$Ni  & 2.33 	& -0.23 & 0.15  & 2.61 		& 2.21 & -0.20 & 0.10&  2.44\tabularnewline
$^{64}$Zn  & 5.22 	& -0.61 & -0.16 & 5.44 		& 0.68 & -0.06 & -0.02 & 0.70\tabularnewline
$^{78}$Kr  & 3.79 	& -0.61 & -0.24 & 3.92 		& 0.87 & -0.14 & -0.06 & 0.90\tabularnewline
$^{96}$Ru  & 2.51 	& -0.37 & 0.11  & 2.85 		& 0.03 & -0.01 & 0.00 & 0.04\tabularnewline
$^{106}$Cd & 3.16 	& -0.38 & 0.19  & 3.59 		& 1.55 & -0.16 & 0.08 & 1.72\tabularnewline
$^{124}$Xe & 4.42 	& -0.82 & -0.19 & 4.74 		& 0.74 & -0.14 & -0.03& 0.80\tabularnewline
$^{130}$Ba & 4.36 	& -0.80 & -0.18 & 4.67 		& 0.32 & -0.06 & -0.01 & 0.34\tabularnewline
$^{136}$Ce & 4.23 	& -0.76 & -0.16 & 4.54 		& 0.35 & -0.06 & -0.01 & 0.38\tabularnewline
$^{156}$Dy & 2.80 	& -0.40 & 0.13 & 3.17 		& 1.53& -0.23 & 0.08 & 1.75\tabularnewline
$^{164}$Er & 3.46 	& -0.44 & 0.22 & 3.95 		& 1.02 & -0.10 & 0.05 & 1.13\tabularnewline
$^{180}$W & 4.12 	& -0.57 & 0.20 & 4.67 		& 0.26 & -0.05 &0.02 & 0.31\tabularnewline

\end{tabular}
\end{table*}
\end{ruledtabular}
The parameters of the IBM-2 Hamiltonian used in this calculation are those in Tables II and VI of Ref.~\cite{barea13b, kotila14}, respectively. Also here, as in the previous Sect. III, we see that the F matrix elements are considerably reduced by isospin restoration in comparison with those without restoration given in Table VIII of Ref.~\cite{barea13b}. This is also seen in Table~\ref{0+chi} where the quantity $\chi_F$ is shown.
\begin{ruledtabular}
\begin{center}
\begin{table}[h]
\caption{\label{0+chi}Ratio Fermi to Gamow-Teller matrix elements, $\chi_F$, for neutrinoless $\beta^{+}\beta^{+}$, $EC\beta^{+}$, and $ECEC$ in the IBM-2 with isospin restoration compared with available QRPA results.}
\begin{tabular}{cccc}
&\multicolumn{2}{c}{$\chi_F$} \\ \cline{2-3}
\T
Decay      		&\multicolumn{2}{c}{IBM-2}  &QRPA\footnotemark[1]\\ \cline{2-3}
&old &new &\\
\T
$^{58}$Ni  &-0.06	& -0.06 &-0.14\tabularnewline
$^{64}$Zn &-0.31	& -0.07 & \tabularnewline
$^{78}$Kr  &-0.38	& -0.10 &-0.27 \tabularnewline
$^{96}$Ru &-0.09	 & -0.09 &-0.23 \tabularnewline
$^{106}$Cd &-0.07	& -0.07 &-0.23 \tabularnewline
$^{124}$Xe  &-0.34	& -0.12 &-0.23 \tabularnewline
$^{130}$Ba  &-0.32	& -0.11 &-0.23 \tabularnewline
$^{136}$Ce &-0.32	& -0.11 &-0.26 \tabularnewline
$^{156}$Dy &		& -0.09 & \tabularnewline
$^{164}$Er &		& -0.08 & \tabularnewline
$^{180}$W &		& -0.09 & \tabularnewline
\end{tabular}
\footnotetext[1]{Reference~\cite{hir94}. No isospin restoration.}
\end{table}
\end{center}
\end{ruledtabular}
Our results are compared with other available calculations in Table~\ref{comp0+}. For $\beta^+\beta^+$, $EC \beta^+$, and $ECEC$ decay there are no QRPA calculations with isospin restoration and thus the comparison is only meant to show the reduction in the F matrix element in IBM-2 brought in by isospin restoration.
\begin{ruledtabular}
\begin{table}[h]
\caption{\label{comp0+}IBM-2 matrix elements with Argonne SRC and isospin restoration for  neutrinoless $\beta^{+}\beta^{+}$, $EC\beta^{+}$, and $ECEC$ compared with available QRPA calculations.}
\begin{tabular}{lccccc}
 &\multicolumn{3}{c}{$0_{1}^{+}$} &\multicolumn{2}{c}{$0_{2}^{+}$}\tabularnewline \cline{2-4} \cline{5-6}
\T
Decay &IBM-2 &\multicolumn{2}{c}{QRPA\footnotemark[1]} &IBM-2 &QRPA\\ 
\hline 
\T
$^{58}$Ni  & 2.61 	&1.55	&	& 2.44 &\tabularnewline
$^{64}$Zn  & 5.44 	&	&	& 0.70 &\tabularnewline
$^{78}$Kr  & 3.92 	&4.16	&	& 0.90 &\tabularnewline
$^{96}$Ru  & 2.85 	&3.23	&4.29\footnotemark[2]	& 0.04 &2.31\footnotemark[2]\tabularnewline
$^{106}$Cd & 3.59 	&4.10	&7.54\footnotemark[3]	& 1.72 &0.61\footnotemark[3]\tabularnewline
$^{124}$Xe & 4.74 	&4.76	&	& 0.80 &\tabularnewline
$^{130}$Ba & 4.67 	&4.95	&	& 0.34 &\tabularnewline
$^{136}$Ce & 4.54 	&3.7	&	& 0.38 &\tabularnewline
$^{156}$Dy & 3.17 &&&1.75&\tabularnewline
$^{164}$Er & 3.95 &&&1.13&\tabularnewline
$^{180}$W & 4.67 &&&0.31&\tabularnewline

\end{tabular}
\footnotetext[1]{Reference~\cite{hir94}. No isospin restoration.}
\footnotetext[2]{Reference~\cite{suh12} (UCOM SRC). No isospin restoration.}
\footnotetext[3]{Reference~\cite{suh11} (UCOM SRC). No isospin restoration.}
\end{table}
\end{ruledtabular}

\subsection{$0\nu\beta^+\beta^+$ and related processes with heavy neutrino exchange}
These matrix elements are obtained in the same way as in Sect. III.2 and are given in Table~\ref{0+heavy}.
\begin{ruledtabular}
\begin{table*}[cbt!]
\caption{\label{0+heavy}Nuclear matrix elements (dimensionless)  for heavy neutrino exchange  for neutrinoless $\beta^{+}\beta^{+}$/$EC\beta^{+}$/$ECEC$ decay in IBM-2 with isospin restoration, Argonne SRC, and $g_V/g_A=1/1.269$.}
\begin{tabular}{lcccccccc}
 & \multicolumn{4}{c}{$0_{1}^{+}$} & \multicolumn{4}{c}{$0_{2}^{+}$}\tabularnewline \cline{2-5} \cline{6-9}
\T
Nucleus& $M_{GT}^{(0\nu)}$ & $M_{F}^{(0\nu)}$  & $M_{T}^{(0\nu)}$ & $M^{(0\nu)}$ & $M_{GT}^{(0\nu)}$ & $M_{F}^{(0\nu)}$  & $M_{T}^{(0\nu)}$ & $M^{(0\nu)}$\tabularnewline
\hline 
\T
$^{58}$Ni  &55.1 	& -23.1 	& 18.6  & 88.0 		& 36.3 & -15.8 	& 8.33 &  54.5\tabularnewline
$^{64}$Zn  & 103 	& -38.9 	& -18.5 & 109 		& 10.1 & -3.20 	& -2.00 & 10.1\tabularnewline
$^{78}$Kr  & 89.8 	& -38.5 	& -30.6 & 83.1 		& 21.1 & -9.12 	& -7.22 & 19.5\tabularnewline
$^{96}$Ru  & 67.5 	& -30.6 	& 12.5  & 99.0 	& 0.32 & -0.08 	& 0.32 & 0.59\tabularnewline
$^{106}$Cd & 87.8 	& -38.1 	& 26.5  & 138 		& 34.0 & -14.7 	& 8.75 & 51.9\tabularnewline
$^{124}$Xe & 105 	& -47.9 	& -25.0 & 110 		& 18.1& -8.24 	& -4.31 & 18.9\tabularnewline
$^{130}$Ba & 103 	& -46.4 	& -23.7 & 108 		& 8.07 & -3.68 	& -1.90 & 8.45\tabularnewline
$^{136}$Ce & 95.8 	& -43.2 	& -21.8 & 101 		& 8.24& -3.73. 	& -1.89 &8.66\tabularnewline
$^{156}$Dy & 82.6 	&-37.0	&17.5&123			&47.6&-21.4	&10.4&71.3\tabularnewline
$^{164}$Er & 108 	&-46.8	&32.9&170			&23.6&-9.95	&5.96&35.8\tabularnewline
$^{180}$W & 119 	&-53.3	&28.1&180			&10.7&-4.85	&2.91&16.6\tabularnewline

\end{tabular}
\end{table*}
\end{ruledtabular}

\subsubsection{Sensitivity to parameter changes, model assumptions and operator
assumptions}

The sensitivity here is identical to that described in Sect. III for $0\nu\beta^-\beta^-$. Our final matrix elements with error estimate are given in  Table~\ref{final0+}.

 \begin{table}[h]
 \caption{\label{final0+}Final $\beta^{+}\beta^{+}$, $EC\beta^{+}$, and $ECEC$ IBM-2 matrix elements with isospin restoration, Argonne SRC, and their error estimate.}
 \begin{ruledtabular} %
\begin{tabular}{ccc}
Decay  & Light neutrino exchange & Heavy neutrino exchange \tabularnewline
\hline 
\T
$^{58}$Ni   	& 2.61(42) 	  & 88(25) 		\tabularnewline
$^{64}$Zn   	& 5.44(87) 	 &109(31) 		\tabularnewline
$^{78}$Kr   	& 3.92(63)	 	& 83(23) 		\tabularnewline
$^{96}$Ru   	& 2.85(46) 	  & 99(28) 	\tabularnewline
$^{106}$Cd  	& 3.59(57) 	  & 138(39) 		\tabularnewline
$^{124}$Xe  	& 4.74(76) 	 & 110(31) 		\tabularnewline
$^{130}$Ba  	& 4.67(75) 	 & 108(30) 		\tabularnewline
$^{136}$Ce  	&4.54(73)  	 & 101(28) 		\tabularnewline
$^{156}$Dy  	&3.17(51)		&123(34)			\tabularnewline
$^{164}$Er  	&3.95(63)		&170(48)			\tabularnewline
$^{180}$W  	&4.67(75)		&180(50)			\tabularnewline
\end{tabular}\end{ruledtabular} 
\end{table}

\section{Results for $2\nu\beta\beta$}
Isospin restoration has a major consequence on matrix elements for $2\nu\beta\beta$ decay, since F matrix elements vanish when isospin restoration is imposed. $2\nu\beta\beta$ matrix elements can be easily calculated in IBM-2 using closure approximation (CA). In this approximation the matrix elements  $M_{2\nu}$, which appear in the half-life Eq. (\ref{hl}) 
can be written as 
\begin{equation}
\begin{split}
M_{2\nu} & =  g_{A}^{2}M^{(2\nu)}, \\
M^{(2\nu)} & =  -\left[\frac{M_{GT}^{(2\nu)}}{\tilde{A}_{GT}}-\left(\frac{g_{V}}{g_{A}}\right)^{2}\frac{M_{F}^{(2\nu)}}{\tilde{A}_{F}}\right],
\end{split}
\end{equation}
where 
\begin{equation}
\begin{split}
M_{GT}^{(2\nu)} & =  \left\langle 0_{F}^{+}\left\vert \sum_{nn^{\prime}}\tau_{n}^{\dag}\tau_{n^{\prime}}^{\dag}\vec{\sigma}_{n}\cdot\vec{\sigma}_{n^{\prime}}\right\vert 0_{I}^{+}\right\rangle , \\
M_{F}^{(2\nu)} & =  \left\langle 0_{F}^{+}\left\vert \sum_{nn^{\prime}}\tau_{n}^{\dag}\tau_{n^{\prime}}^{\dag}\right\vert 0_{I}^{+}\right\rangle .
\end{split}
\end{equation}
The closure energies $\tilde{A}_{GT}$ and $\tilde{A}_{F}$ are defined
by
\begin{equation}
\begin{split}
\tilde{A}_{GT} & =  \frac{1}{2}\left(Q_{\beta\beta}+2m_{e}c^{2}\right)+\left\langle E_{1^{+},N}\right\rangle -E_{I}, \\
\tilde{A}_{F} & =  \frac{1}{2}\left(Q_{\beta\beta}+2m_{e}c^{2}\right)+\left\langle E_{0^{+},N}\right\rangle -E_{I},
\end{split}
\end{equation}
where $\left\langle E_{N}\right\rangle $ is a suitably chosen excitation
energy in the intermediate odd-odd nucleus. The matrix elements $M^{(2\nu)}$ can be simply calculated by replacing the neutrino potential $v(p)$
\begin{equation}
v_{2\nu}(p)=\frac{\delta(p)}{p^2},
\end{equation}
which is the Fourier-Bessel transform of the configuration space potential $V(r)=1$.

In order to confirm that in isospin-restored IBM-2 calculation the Fermi matrix elements for $2\nu\beta\beta$ decay vanish we have calculated $M^{(2\nu)}_{GT}$ and $M^{(2\nu)}_{F}$. The results are given in Table~\ref{table7}.
\begin{table}[cbt!]
 \caption{\label{table7}$2\nu\beta^-\beta^-$ matrix elements (dimensionless) to the ground state
(columns 2 and 3) and to the first excited state (columns 4 and 5)
using the microscopic interacting boson model (IBM-2) with  isospin restoration and and Argonne SRC in the closure
approximation.}
 \begin{ruledtabular} %
\begin{tabular}{ccccc}
&\multicolumn{2}{c}{$0_{1}^{+}$} &\multicolumn{2}{c}{$0_{2}^{+}$}\\\cline{2-3}\cline{4-5}
\T
Nucleus  & $M_{GT}^{(2\nu)}$  & $M_{F}^{(2\nu)}$   & $M_{GT}^{(2\nu)}$  & $M_{F}^{(2\nu)}$ \tabularnewline
\hline 
\T
$^{48}$Ca  &1.64	&-0.01		&5.07	&-0.01\tabularnewline
$^{76}$Ge  &4.44	&-0.01		&2.02	&-0.00 \tabularnewline
$^{82}$Se  &3.59	&-0.01		&1.05	&-0.00 \tabularnewline
$^{96}$Zr  &2.28	&-0.00		&0.04	&-0.00 \tabularnewline
$^{100}$Mo  &3.05	&-0.00		&0.81	&-0.00 \tabularnewline
$^{110}$Pd  &3.08	&-0.00		&0.38	&-0.00 \tabularnewline
$^{116}$Cd  &2.38	&-0.00		&0.83	&-0.00 \tabularnewline
$^{124}$Sn  &2.86	&-0.01		&2.19	&-0.00\tabularnewline
$^{128}$Te  &3.71	&-0.01		&2.70	&-0.00 \tabularnewline
$^{130}$Te &3.39	&-0.01		&2.64	&-0.00 \tabularnewline
$^{134}$Xe &3.69  	&-0.01        	&2.34       &-0.00 \tabularnewline
$^{136}$Xe &2.82	&-0.01		&1.65	&-0.00 \tabularnewline
$^{148}$Nd  &1.31	&-0.00		&0.18	&-0.00 \tabularnewline
$^{150}$Nd  &1.61	&-0.00		&0.31	&-0.00 \tabularnewline
$^{154}$Sm  &1.95	&-0.00		&0.35	&-0.00 \tabularnewline
$^{160}$Gd  &3.08	&-0.00		&0.53	&-0.00 \tabularnewline
$^{198}$Pt  &1.06	&-0.00		&0.03	&-0.00 \tabularnewline
$^{232}$Th  &2.75	&-0.00		&0.08	&-0.00 \tabularnewline
$^{238}$U  &3.35	&-0.00		&0.24	&-0.00 \tabularnewline

\end{tabular}
\end{ruledtabular} 
\end{table}
We can see from this table that $M_F^{(2\nu)}$ indeed vanish. The small values $\sim 0.01$ are an indication of our numerical accuracy in calculating the radial integrals $R^{k_1,k_2,\lambda}$.

\begin{figure}[h]
\begin{center}
\includegraphics[width=8.6cm]{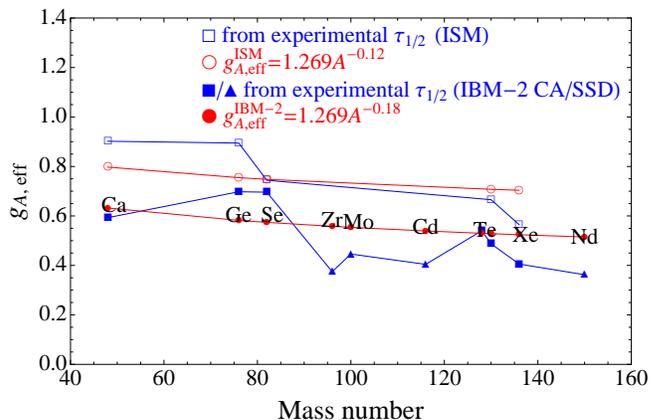} 
\end{center}
\caption{\label{fig5}(Color online) Value of $g_{A,eff}$ extracted from experiment for IBM-2 and ISM.}
\end{figure}

Using the results in Table~\ref{table7} one can redo the analysis of Ref.~\cite{kotila} and extract the values of the effective $g_{A,eff}$ from 
\begin{equation}
M_{2\nu}^{eff}=\left( \frac{g_{A,eff}}{g_A}\right)^2M_{2\nu}
\end{equation}
with $|M_{2\nu}^{eff}|$ extracted from experiment \cite{barabash} as compiled in Ref. \cite{kotila}. The corresponding results are shown in Fig.~\ref{fig5}. Isospin restoration has no effect on the extracted values of $g_{A,eff}$, since in the previous analysis \cite{barea13} the Fermi matrix elements $M_F^{(2\nu)}$ were simply discarded. The difference between Fig. 5 of this paper and Fig. 13 of \cite{barea13} is only due to the fact that we have used Argonne SRC instead of Miller-Spencer. 

We also note that,  very recently, $g_{A,eff}$ values have also been extracted in QRPA-T\"u \cite{eng14} and QRPA-Jy \cite{suh13} with results similar to those in Fig.~\ref{fig5}.

\section{Results for $2\nu\beta^+\beta^+$ and competing modes}
Matrix elements for $2\nu\beta^+\beta^+$ and related processes can be obtained in the same way as in Sect. V. The results are given in Table~\ref{2+}.
\begin{ruledtabular}
\begin{table}[cbt!]
\caption{\label{2+}$2\nu\beta^{+}\beta^{+}$, $2\nu EC\beta^{+}$, and $2\nu ECEC$ nuclear matrix elements (dimensionless) to the ground state
(columns 2 and 3) and to the first excited state (columns 4 and 5)
using IBM-2 with  isospin restoration and and Argonne SRC in the closure
approximation.}
\begin{tabular}{ccccc}
&\multicolumn{2}{c}{$0_{1}^{+}$} &\multicolumn{2}{c}{$0_{2}^{+}$}\\\cline{2-3}\cline{4-5}
\T
Nucleus  & $M_{GT}^{(2\nu)}$  & $M_{F}^{(2\nu)}$   & $M_{GT}^{(2\nu)}$  & $M_{F}^{(2\nu)}$  \tabularnewline
\hline 
\T
$^{58}$Ni  &2.11	&-0.00		&2.34	&-0.00		\tabularnewline
$^{64}$Zn  &5.20	&-0.01		&0.71	&-0.00		\tabularnewline
$^{78}$Kr  &3.67	&-0.01		&0.83	&-0.00		\tabularnewline
$^{96}$Ru  &2.17	&-0.00		&0.05	&-0.00		\tabularnewline
$^{106}$Cd &2.57	&-0.00		&1.47	&-0.00		\tabularnewline
$^{124}$Xe &4.24	&-0.01		&0.71	&-0.00		\tabularnewline
$^{130}$Ba &4.22	&-0.01		&0.30	&-0.00		\tabularnewline
$^{136}$Ce &4.17	&-0.01		&0.34	&-0.00		\tabularnewline
$^{156}$Dy &2.20	&-0.00		&1.15	&-0.00		\tabularnewline
$^{164}$Er &2.58	&-0.00		&0.90	&-0.00		\tabularnewline
$^{180}$W &3.09	&-0.01		&0.12	&-0.00		\tabularnewline

\end{tabular}
\end{table}
\end{ruledtabular}

\section{Expected half-lives and limits on neutrino mass}
\subsection{Light neutrino exchange}
\begin{figure}[h]
\begin{center}
\includegraphics[width=8.6cm]{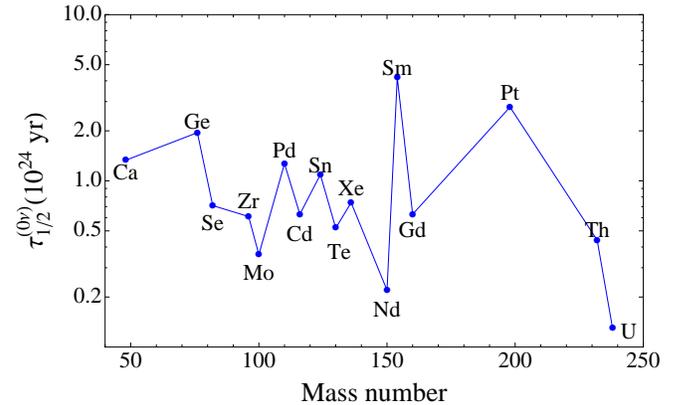} 
\end{center}
\caption{\label{fig6}(Color online) Expected half-lives for $\left\langle m_{\nu}\right\rangle=1$~eV, $g_{A}=1.269$ and IBM-2 isospin restored nuclear matrix elements. The points for $^{128}$Te, $^{134}$Xe, and $^{148}$Nd decays are not included in this figure. The figure is in semilogarithmic scale.}
\end{figure}

The calculation  of nuclear matrix elements in IBM-2 with isospin restoration can now be combined with phase-space factors \cite{kotila,kotila14,kotila13} to produce our final results for half-lives for light neutrino exchange in Table~\ref{table14} and Fig.~\ref{fig6}.

\begin{ruledtabular}
\begin{table}[h]
\caption{\label{table14}Left: Calculated half-lives in IBM-2 Argonne SRC for neutrinoless double-$\beta$ decay for $\left<m_{\nu}\right>=1$~eV and $g_A=1.269$. Right: Upper limit on neutrino mass from current experimental limit from a compilation of Barabash \cite{barabash11}. The value reported by Klapdor-Kleingrothaus \textit{et al.} \cite{klapdor}, IGEX collaboration \cite{igex}, and the recent limits from KamLAND-Zen \cite{kamland}, EXO \cite{exo0nu}, and GERDA \cite{gerda} are also included.}
\begin{tabular}{lc|cc}
Decay  &  \ensuremath{\tau_{1/2}^{0\nu}}(\ensuremath{10^{24}}yr) &  \ensuremath{\tau_{1/2, exp}^{0\nu}}(yr) &$\left< m_{\nu}\right>$ (eV)\\
 \hline
 \T
$^{48}$Ca$\rightarrow ^{48}$Ti		&1.33 &$>5.8\times 10^{22}$ &$<4.8$\\
$^{76}$Ge$\rightarrow ^{76}$Se 	&1.95 &$>1.9\times 10^{25}$ &$<0.32$\\
							&	 	&$1.2\times 10^{25}$\footnotemark[1] &$0.40$\\
							&	 	&$>1.6\times 10^{25}$\footnotemark[2] &$<0.35$\\
							&	 	&$>2.1\times 10^{25}$\footnotemark[3] &$<0.30$\\
$^{82}$Se$\rightarrow ^{82}$Kr	 	&0.71 &$>3.6\times 10^{23}$ &$<1.4$\\
$^{96}$Zr$\rightarrow ^{96}$Mo	&0.61 &$>9.2\times 10^{21}$ &$<8.1$\\
$^{100}$Mo$\rightarrow ^{100}$Ru 	&0.36 &$>1.1\times 10^{24}$ &$<0.57$\\
$^{110}$Pd$\rightarrow ^{110}$Cd 	&1.27 & &\\
$^{116}$Cd$\rightarrow ^{116}$Sn  	&0.63 &$>1.7\times 10^{23}$ &$<1.9$\\
$^{124}$Sn$\rightarrow ^{124}$Te 	&1.09 & &\\
$^{128}$Te$\rightarrow ^{128}$Xe 	&10.19 &$>1.5\times 10^{24}$ &$<2.6$\\
$^{130}$Te$\rightarrow ^{130}$Xe 	&0.52 &$>2.8\times 10^{24}$ &$<0.43$\\
$^{134}$Xe$\rightarrow ^{124}$Ba 	&10.23 & &\\
$^{136}$Xe$\rightarrow ^{136}$Ba 	&0.74 &$>1.9\times 10^{25}$\footnotemark[4] &$<0.20$\\
								 	&	 &$>1.6\times 10^{25}$\footnotemark[5] &$<0.22$\\

$^{148}$Nd$\rightarrow ^{148}$Sm 	&1.87 & &\\
$^{150}$Nd$\rightarrow ^{150}$Sm 	&0.22 &$>1.8\times 10^{22}$ &$<3.5$\\
$^{154}$Sm$\rightarrow ^{154}$Gd 	&4.19 & &\\
$^{160}$Gd$\rightarrow ^{160}$Dy 	&0.63 & &\\
$^{198}$Pt$\rightarrow ^{198}$Hg 	&2.77 & &\\
$^{232}$Th$\rightarrow ^{232}$U 	&0.44\\
$^{238}$U$\rightarrow ^{238}$Pu 	&0.13\\
\end{tabular}
\footnotetext[1]{Ref.~\cite{klapdor}}
\footnotetext[2]{Ref.~\cite{igex}}
\footnotetext[3]{Ref.~\cite{gerda}}
\footnotetext[4]{Ref.~\cite{kamland}}
\footnotetext[5]{Ref.~\cite{exo0nu}}

\end{table}
\end{ruledtabular}

\begin{figure}[cbt!]
\begin{center}
\includegraphics[width=8.6cm]{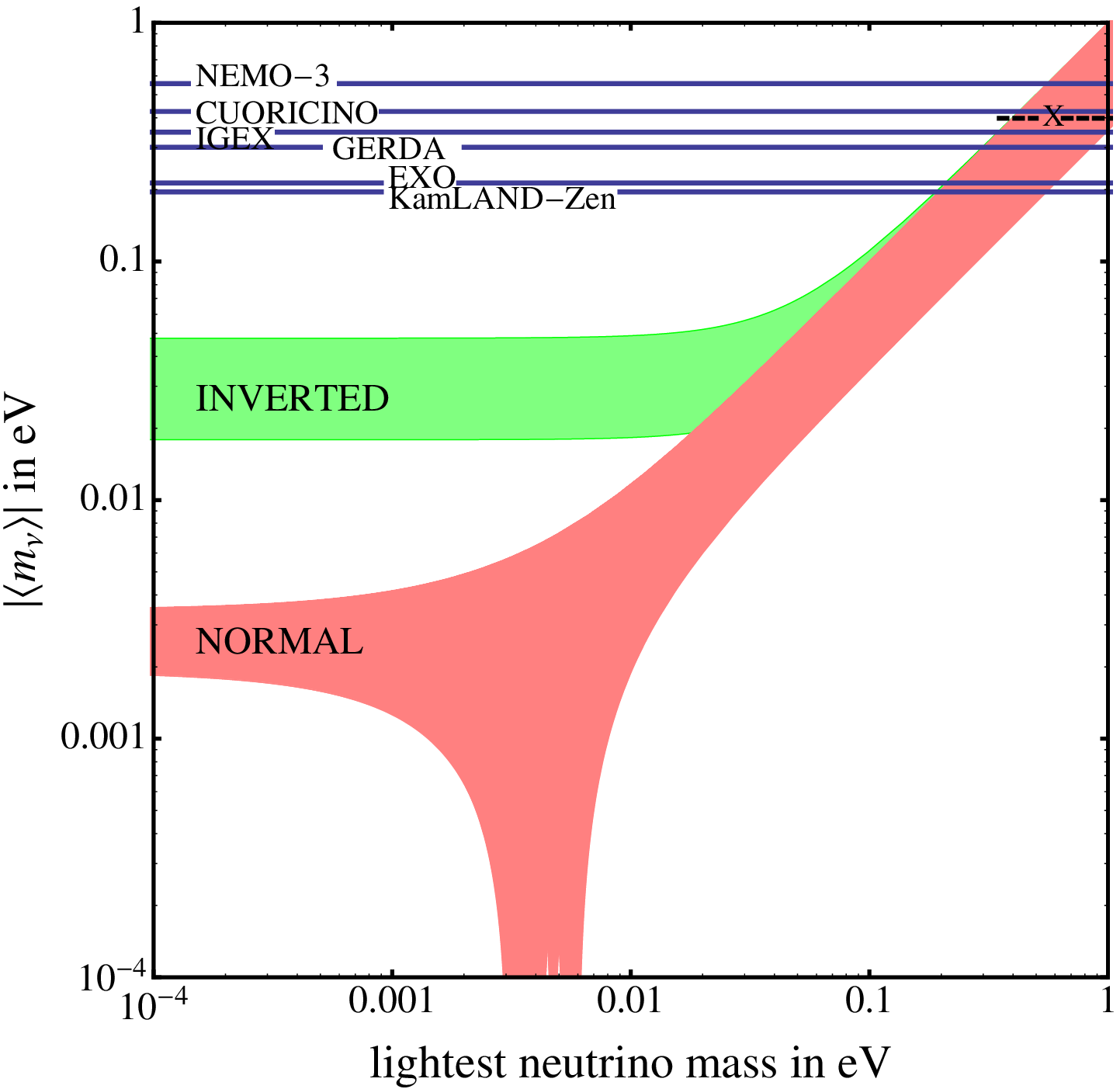} 
\end{center}
\caption{\label{fig7}(Color online) Current limits to $\left\langle m_{\nu}\right\rangle$ from CUORICINO \cite{cuoricino}, IGEX \cite{igex}, NEMO-3 \cite{nemo}, KamLAND-Zen \cite{kamland}, EXO \cite{exo0nu}, and GERDA \cite{gerda}, and IBM-2 Argonne SRC isospin restored nuclear matrix elements and $g_A=1.269$. The value of Ref.~\cite{klapdor} is shown by $X$. The figure is in logarithmic scale.}
\end{figure}

For light neutrino exchange 
\begin{equation}
\left\vert f(m_{i},U_{ei})\right\vert ^{2}=\left\vert \frac{\langle m_\nu\rangle}{m_e}\right\vert ^{2}.
\end{equation}
The average light neutrino mass is constrained by atmospheric, solar,
reactor and accelerator neutrino oscillation experiments to be \cite{fogli}
\begin{equation}
\begin{split}
\left\langle m_{\nu}\right\rangle   =  \left\vert c_{13}^{2}c_{12}^{2}m_{1}\right.&\left.+c_{13}^{2}s_{12}^{2}m_{2}e^{i\varphi_{2}}+s_{13}^{2}m_{3}e^{i\varphi_{3}}\right\vert , \\
c_{ij}  =  \cos\vartheta_{ij},\text{ \ \ }&s_{ij}=\sin\vartheta_{ij},\text{ \ \ }\varphi_{2,3}=[0,2\pi],\\
\left(m_{1}^{2},m_{2}^{2},m_{3}^{2}\right)  =&
 \frac{m_{1}^{2}+m_{2}^{2}}{2}
+\left(-\frac{\delta m^{2}}{2},+\frac{\delta m^{2}}{2},\pm\Delta m^{2}\right).
\end{split}
\end{equation}
Using the best fit values \cite{fogli} 
\begin{equation}
\begin{split}
\sin^{2}\vartheta_{12} &=0.213,\text{\ \ }\sin^{2}\vartheta_{13}=0.016,\\
\sin^{2}\vartheta_{23}&=0.466,\text{\ \ }\delta m^{2} =7.67\times10^{-5}\text{~eV}^{2},\\
\Delta m^{2}&=2.39\times10^{-3}\text{~eV}^{2}
\end{split}
\end{equation}
we obtain the values given in Fig.~\ref{fig7}. In this figure we have added
the current limits, for $g_{A}=1.269$, coming from CUORICINO \cite{cuoricino}, IGEX \cite{igex}, NEMO-3 \cite{nemo}, KamLAND-Zen \cite{kamland}, EXO \cite{exo0nu} and GERDA \cite{gerda} experiments. Also, henceforth we use  c=1 to conform with standard notation.

\subsection{Heavy neutrino exchange}
The half-lives for this case are calculated using the formula 
\begin{equation}
\begin{split}
\lbrack\tau_{1/2}^{0\nu_{h}}]^{-1} & =  G_{0\nu}^{(0)}\left\vert M_{0\nu_{h}}\right\vert ^{2}\left\vert \eta\right\vert ^{2}\\\
\eta & \equiv  m_{p}\left\langle m_{\nu_{h}}^{-1}\right\rangle =\sum_{k=heavy}\left(U_{ek_h}\right)^{2}\frac{m_{p}}{m_{k_h}}.
\end{split}
\end{equation}
The expected half-lives for $|\eta|=10^{-7}$, and using the IBM-2 matrix elements of Table~\ref{table4},
are shown in Table~\ref{table15}. For other values of $\eta$ they scale as $|\eta|^2$.
There are no direct experimental bounds on $\eta $.
Recently, Tello\textit{\ et al.} \cite{tello} have argued that from
lepton flavor violating processes and from large hadron collider (LHC) experiments one can
put some bounds on the right-handed leptonic mixing matrix $U_{ek,heavy}$
and thus on $\eta$. In the model of Ref.~\cite{tello}, when converted to our notation, $\eta$ can be written as 
\begin{equation}
\eta=\frac{M_W^4}{M_{WR}^4}\sum_{k=heavy}\left( V_{ek_h}\right)^2\frac{m_p}{m_{k_h}},
\end{equation}
where $M_W$ is the mass of the $W$-boson, $M_W=(80.41\pm 0.10)$~GeV \cite{experiment}, $M_{WR}$ is the mass of $WR$-boson, and $V=\left( M_{WR}/M_W \right)^2 U$.  By comparing the calculated half-lives with their current experimental limits, we can set limits on the lepton nonconserving parameter $|\eta|$, shown in Table~\ref{table15}.

If we write
\begin{equation}
\eta=\frac{M_W^4}{M_{WR}^4} \frac{m_p}{ \langle m_{\nu_h} \rangle },
\end{equation}
we can also set limits on the average heavy neutrino mass, $\langle m_{\nu_h} \rangle$. This limit is model dependent. In Ref.~\cite{tello} a value of $M_{WR}=3.5$~TeV was used. We use here instead a lower value  $M_{WR}=1.75$~TeV, obtaining the limits on $\langle m_{\nu_h} \rangle$ shown in the last column of Table~\ref{table15}. For other values of $M_{WR}$ it scales as $M_{WR}^{-4}$.
\begin{ruledtabular}
\begin{center}
\begin{table*}[cbt!]
\caption{\label{table15}Left: Calculated half-lives for neutrinoless double-$\beta$ decay with exchange of heavy neutrinos for $\eta=2.75\times10^{-7}$ and $g_A=1.269$. Right: Upper limits of $|\eta|$ and lower limits of heavy neutrino mass (see text for details) from current experimental limit from a compilation of Barabash \cite{barabash11}. The value reported by Klapdor-Kleingrothaus \textit{et al.} \cite{klapdor}, IGEX collaboration \cite{igex}, and the recent limit from KamLAND-Zen \cite{kamland}, EXO \cite{exo0nu}, and GERDA \cite{gerda} are also included.}
\begin{tabular}{lc|ccc}
Decay  &  \ensuremath{\tau_{1/2}^{0\nu_h}}(\ensuremath{10^{24}}yr) &  \ensuremath{\tau_{1/2, exp}^{0\nu_h}}(yr) &$|\eta|(10^{-6})$ &$\left< m_{\nu_h}\right>$(GeV)\\
 \hline
 \T
$^{48}$Ca$\rightarrow ^{48}$Ti		&0.72	&$>5.8\times 10^{22}$				&<0.36		&>11.9\\
$^{76}$Ge$\rightarrow ^{76}$Se	 &1.51	&$>1.9\times 10^{25}$				&<0.028		&>148\\
							&	 	&$1.2\times 10^{25}$\footnotemark[1]&0.035 	&118\\
						 	&	 	&$>1.6\times 10^{25}$\footnotemark[2]&<0.031 	&>136\\
							&	 	&$>2.1\times 10^{25}$\footnotemark[3] &$<0.027$	&156\\

$^{82}$Se$\rightarrow ^{82}$Kr	 	&0.55	&$>3.6\times 10^{23}$				&<0.12		&>34\\
$^{96}$Zr$\rightarrow ^{96}$Mo	&0.19	&$>9.2\times 10^{21}$				&<0.46		&>9.15\\
$^{100}$Mo$\rightarrow ^{100}$Ru 	&0.09	&$>1.1\times 10^{24}$				&<0.028	&>146\\
$^{110}$Pd$\rightarrow ^{110}$Cd 	&0.33	&									&			&\\
$^{116}$Cd$\rightarrow ^{116}$Sn  	&0.19	&$>1.7\times 10^{23}$				&<0.11		&>39.5\\
$^{124}$Sn$\rightarrow ^{124}$Te 	&0.67	&									&			&\\
$^{128}$Te$\rightarrow ^{128}$Xe 	&6.43	&$>1.5\times 10^{24}$				&<0.21		&>20.2\\
$^{130}$Te$\rightarrow ^{130}$Xe 	&0.32	&$>2.8\times 10^{24}$				&<0.034		&>123\\
$^{134}$Xe$\rightarrow ^{134}$Ba 	&8.57	&	&	&\\
$^{136}$Xe$\rightarrow ^{136}$Ba 	&0.50	&$>1.9\times 10^{25}$\footnotemark[4]&<0.016	&>257\\
								 	&		&$>1.6\times 10^{25}$\footnotemark[5]&<0.018	&>236\\
$^{148}$Nd$\rightarrow ^{148}$Sm 	&0.36	&									&			&\\
$^{150}$Nd$\rightarrow ^{150}$Sm 	&0.05	&$>1.8\times 10^{22}$				&<0.16		&>26.3\\
$^{154}$Sm$\rightarrow ^{154}$Gd 	&1.00	&									&			&\\
$^{160}$Gd$\rightarrow ^{160}$Dy 	&0.17	&									&			&\\
$^{198}$Pt$\rightarrow ^{198}$Hg 	&0.48	&									&			&\\
$^{232}$Th$\rightarrow ^{232}$U 	&0.11\\
$^{238}$U$\rightarrow ^{238}$Pu 	&0.03\\
\end{tabular}
\footnotetext[1]{Ref.~\cite{klapdor}}
\footnotetext[2]{Ref.~\cite{igex}}
\footnotetext[3]{Ref.~\cite{gerda}}
\footnotetext[4]{Ref.~\cite{kamland}}
\footnotetext[5]{Ref.~\cite{exo0nu}}
\end{table*}
\end{center}
\end{ruledtabular}

If both light and heavy neutrino exchange contribute, the half-lives are given by 
\begin{equation}
\lbrack\tau_{1/2}^{0\nu}]^{-1} = G_{0\nu}^{(0)}\left\vert M_{0\nu} \frac{\langle m_{\nu}\rangle }{m_e}  +  M_{0\nu_{h}} \eta \right\vert ^{2}.
\end{equation}

\subsection{Expected half-lives for double positron decay}
\begin{figure}[h]
\begin{center}
\includegraphics[width=8.6cm]{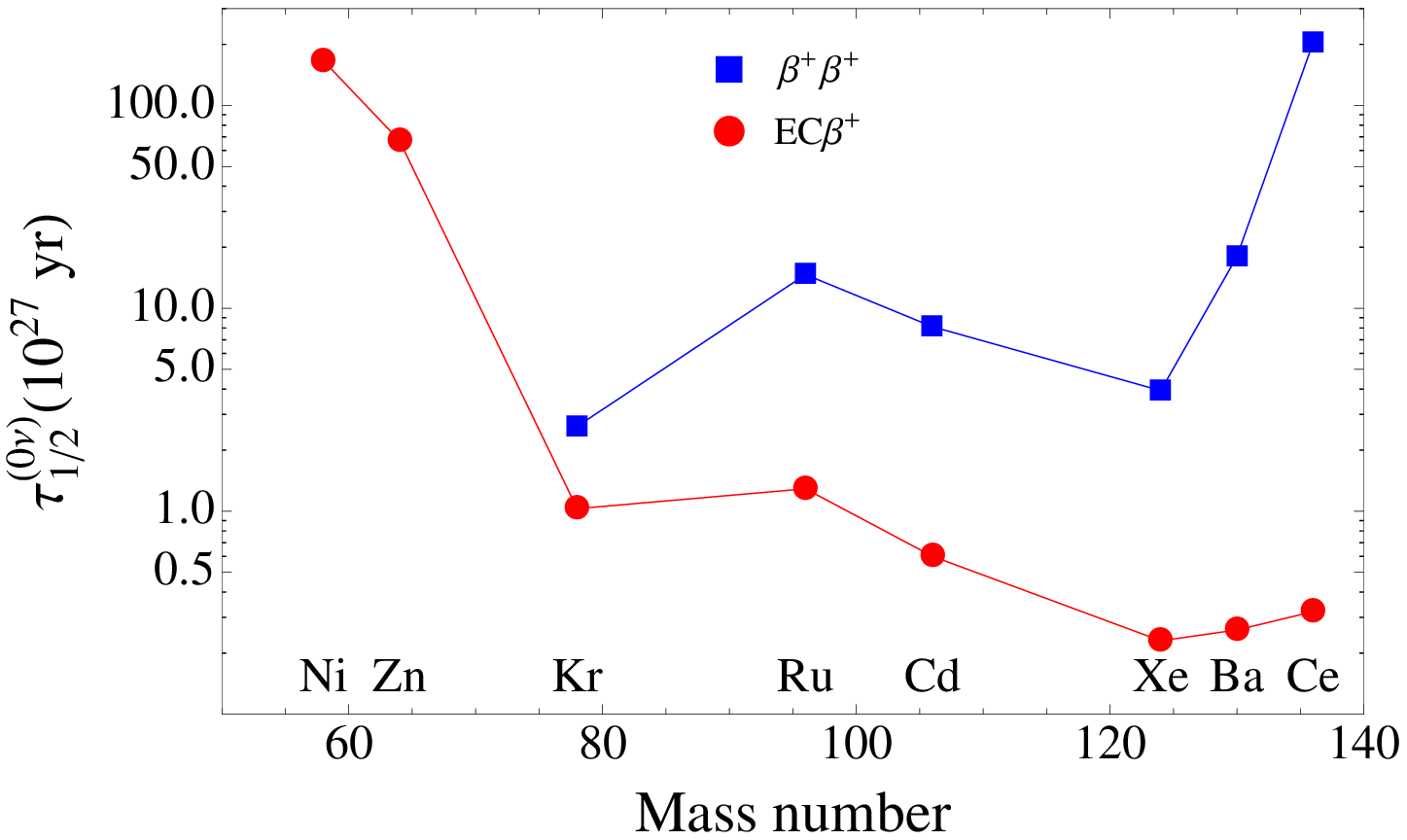} 
\end{center}
\caption{\label{fig8}(Color online) Expected half-lives for $\left\langle m_{\nu}\right\rangle=1$~eV, $g_{A}=1.269$ and IBM-2 isospin restored nuclear matrix elements.  The figure is in semilogarithmic scale.}
\end{figure}

Although no limits are available here, we include for completeness in Fig.~\ref{fig8} our expected half-lives for double positron decay and positron emitting electron capture.

The matrix elements reported in Ref.~\cite{kotila14} for $R0\nu ECEC$ were already obtained with isospin restoration  and therefore expected half-lives for this process are not reported here.

\subsection{Effect of renormalization of $g_A$ on expected half-lives}
Half-lives of $0\nu\beta\beta$ decay depend on $g_A$ as $g_A^4$. In Fig. 5 we have shown the values of $g_{A,eff}$ both for IBM-2 with isospin restoration and for ISM as extracted from $2\nu\beta\beta$ decay. There is much discussion at the present time on whether or not $0\nu\beta\beta$ is equally quenched as $2\nu\beta\beta$ (and single $\beta$-decay). In order to investigate the possible impact of quenching of $g_A$ we present in Table~\ref{tablegaeff} the predicted half-lives under the assumption of maximal quenching 
\begin{equation}
\begin{split}
g_{A,eff}^{\rm{IBM-2}}&=1.269A^{-0.18},\\
g_{A,eff}^{\rm{ISM}}&=1.269A^{-0.12}
\end{split}
\end{equation}
and compare them with the unquenched values with $g_A=1.269$ also given in Table~\ref{tablegaeff}. Quenching of $g_A$ appears to have a major effect on the calculated half-lives, by multiplying them by a factor of 6-50.The question of what value of $g_A$ one needs to use is thus a major concern  which needs to be addressed. This concern has also been discussed recently in Refs. \cite{rob13,del14}.

\begin{ruledtabular}
\begin{center}
\begin{table*}[cbt!]
\caption{\label{tablegaeff}Predicted half-lives in $0\nu\beta\beta$ decay with unquenched and maximally quenched $g_A$,  $g_{A,eff}^{\rm{IBM-2}}$ and $g_{A,eff}^{\rm{ISM}}$ obtained from $2\nu\beta\beta$ decay and $\langle m_\nu\rangle=1$eV.}
\begin{tabular}{lcccc}
&\multicolumn{4}{c}{\ensuremath{\tau_{1/2}^{0\nu}}(\ensuremath{10^{24}}yr)}\\ \cline{2-5}
\T&\multicolumn{2}{c}{IBM-2} &\multicolumn{2}{c}{ISM}\\ \cline{2-3} \cline{4-5}
\T
Decay  &unquenched &maximally quenched	&unquenched &maximally quenched\\
 \hline
 \T
$^{48}$Ca$\rightarrow ^{48}$Ti		&1.33	&21.5	&13.9	&89.2\\
$^{76}$Ge$\rightarrow ^{76}$Se	&1.95 	&44.0	&8.65	&69.1\\
$^{82}$Se$\rightarrow ^{82}$Kr		&0.71 	&16.9	&2.22	&18.5\\
$^{96}$Zr$\rightarrow ^{96}$Mo	&0.61	&16.3		&&\\
$^{100}$Mo$\rightarrow ^{100}$Ru	&0.36 	&9.8		&&\\
$^{110}$Pd$\rightarrow ^{110}$Cd	&1.27 	&37.6		&&\\
$^{116}$Cd$\rightarrow ^{116}$Sn 	&0.63	&19.2		&&\\
$^{124}$Sn$\rightarrow ^{124}$Te 	&1.09	&35.2	&2.73	&27.6\\
$^{128}$Te$\rightarrow ^{128}$Xe 	&10.19	&335		&33.5	&344.4\\
$^{130}$Te$\rightarrow ^{130}$Xe 	&0.52	&17.2	&1.70	&17.6\\
$^{134}$Xe$\rightarrow ^{134}$Ba 	&10.23	&348	&	&\\

$^{136}$Xe$\rightarrow ^{136}$Ba 	&0.74	&25.5	&2.39	&25.3\\
$^{148}$Nd$\rightarrow ^{148}$Sm 	&1.87	&68.2		&&\\
$^{150}$Nd$\rightarrow ^{150}$Sm 	&0.22	&8.3		&&\\
$^{154}$Sm$\rightarrow ^{154}$Gd 	&4.19	&158		&&\\
$^{160}$Gd$\rightarrow ^{160}$Dy 	&0.63	&24.4		&&\\
$^{198}$Pt$\rightarrow ^{198}$Hg 	&2.77	&125	&&\\
$^{232}$Th$\rightarrow ^{232}$U 	&0.44	&22.4		&&\\
$^{238}$U$\rightarrow ^{238}$Pu 	&0.13	&6.7		&&\\
\end{tabular}
\end{table*}
\end{center}
\end{ruledtabular}

\section{Conclusions}

In this paper, we have introduced a method for restoration of the isospin properties of the Fermi transition operator in the calculation of the Fermi matrix elements within of the framework of IBM-2, and done a consistent calculation of $0\nu\beta\beta$, $0\nu_h\beta\beta$, and $2\nu\beta\beta$ NME in the closure approximation. With this method, the F matrix elements for $2\nu\beta\beta$ decay are set to zero,  and those for $0\nu\beta\beta$ and $0\nu_h\beta\beta$ are smaller, with $\chi_F$ factors of order $\sim 0.10$ for all nuclei. 

\acknowledgements
This work was supported in part by US Department of Energy (Grant No. DE-FG-02-91ER-40608), Chilean
Ministry of Education Fondecyt (Grant No. 1120462), and Academy of Finland (Project 266437).


\begin{thebibliography}{100}
\bibitem{cremonesi} O. Cremonesi and M. Pavan, Adv. High En. Phys. \textbf{2014}, 951432 (2014).
\bibitem{kotila} J. Kotila and F.\ Iachello, Phys. Rev. C \textbf{85},
034316 (2012).
\bibitem{barea} J.\ Barea and F.\ Iachello, Phys. Rev. C \textbf{79},
044301 (2009).
\bibitem{barea12} J.\ Barea, J. Kotila, and F.\ Iachello, Phys. Rev. Lett. \textbf{109},
042501 (2012).
\bibitem{barea13} J.\ Barea, J. Kotila and F.\ Iachello, Phys. Rev. C \textbf{87},
014315 (2013).
\bibitem{barea13b} J.\ Barea, J. Kotila and F.\ Iachello, Phys. Rev. C \textbf{87},
057301 (2013).
\bibitem{kotila14} J. Kotila, J. Barea, and F.\ Iachello, Phys. Rev. C  \textbf{89}, 064319 (2014).
\bibitem{kotila13} J. Kotila and F.\ Iachello, Phys. Rev. C \textbf{87},
024313 (2013).
\bibitem{x} F. \v{S}imkovic,  A. Faessler, V.\ Rodin, P. Vogel and
J. Engel, Phys.\ Rev. C \textbf{77}, 045503 (2008).
\bibitem{xx} J. Suhonen, J. Phys. G \textbf{19}, 139 (1993).
\bibitem{sim13} F. \v{S}imkovic, V.\ Rodin, A. Faessler, and
P. Vogel, Phys.\ Rev. C \textbf{87}, 045501 (2013).
\bibitem{ell80} J. P. Elliott and A. P. White, Phys. Lett. B  \textbf{97}, 169 (1980).
\bibitem{ell81} J. P. Elliott and J. A. Evans, Phys. Lett. B \textbf{101}, 216 (1981).
\bibitem{ell87} J. P. Elliott and J. A. Evans, Phys. Lett. B \textbf{195}, 1 (1987).
\bibitem{ell87b} J. P. Elliott, J. A. Evans, and A. P. Williams, Nucl. Phys. A \textbf{469}, 51  (1987).
\bibitem{hal84} P. Halse, J. P. Elliott, and J. A. Evans, Nucl. Phys. A \textbf{417}, 301 (1984).
\bibitem{ell90} J. P. Elliott, Prog. Part. Nucl. Phys.  \textbf{25}, 325 (1990), and references therein.
\bibitem{bel13}J. Beller \textit{et al.}, Phys. Rev. Lett.  \textbf{111},  172501 (2013).
\bibitem{caurier2007} E.\ Caurier, F. Nowacki, and A.\ Poves, Int. J. Mod. Phys. E \textbf{16}, 552 (2007).
\bibitem{poves} J.\ Men\'{e}ndez, A.\ Poves, E. Caurier, and F. Nowacki, Nucl. Phys. A. \textbf{818}, 139 (2009).
\bibitem{fae14} A. Faessler, M. Gonz\'alez, S. Kovalenko, and F. \v{S}imkovic,  
 Phys.\ Rev. D \textbf{90}, 096010 (2014).
\bibitem{nea14} A. Neacsu, S. Stoica, Adv. High En. Phys. \textbf{2014}, 724315 (2014).
\bibitem{Martinez-Pinedo} T. R. Rodr\'{i}guez and G.\ Mart\'{i}nez-Pinedo,
Phys. Rev.\ Lett. \textbf{105}, 252503 (2010).
\bibitem{chandra} K. Chaturvedi, R. Chandra, P.K.\ Rath, P.K.\ Raina,
and J.G.\ Hirsch, Phys. Rev. C \textbf{78}, 054302 (2008).
\bibitem{suh} J. Suhonen, \textit{Proc. of Beauty in Physics: Theory and Experiment},
AIP Conf.\ Proc. \textbf{1488}, 326 (2012), and references therein.
\bibitem{suhpriv}J. Suhonen, private communication .
\bibitem{fan11} D.-L. Fang, A. Faessler, V. Rodin, and F. \v{S}imkovic, Phys. Rev. C \textbf{83}, 034320 (2011).
\bibitem{hir94} M. Hirsch, K. Muto, T. Oda, and H. V. Klapdor-Kleingrothaus, Z. Phys. A \textbf{347}, 151 (1994).
\bibitem{suh12}J. Suhonen, Phys. Rev. C \textbf{86}, 024301 (2012).
\bibitem{suh11}J. Suhonen, Phys. Lett. B \textbf{701}, 490 (2011).
\bibitem{barabash} A.S.\ Barabash, Phys. Rev. C \textbf{81}, 035501(2010). 
\bibitem{eng14}J. Engel, F. \v{S}imkovic, and P Vogel, Phys. Rev. C \textbf{89}, 064308 (2014).
\bibitem{suh13}J. Suhonen and O. Civitarese, Phys. Lett. B \textbf{725}, 153 (2013).
\bibitem{fogli} G. L. Fogli \textit{et al.}, Phys. Rev. D \textbf{75},
053001 (2007); D \textbf{78}, 033010 (2008).
\bibitem{cuoricino}C. Arnaboldi \textit{et al.} (CUORICINO collaboration), Phys. Rev. C \textbf{78}, 035502 (2008).
\bibitem{igex}C. E. Aalseth \textit{et al.} (IGEX collaboration), Phys. Rev. D \textbf{65}, 092007 (2002).
\bibitem{nemo} R. Arnold, \textit{et al.} (NEMO collaboration), Nucl. Phys. A \textbf{765}, 483 (2006).

\bibitem{kamland} A. Gando \textit{et al.} (KamLAND-Zen collaboration), Phys. Rev. Lett. \textbf{110}, 062502 (2013)
\bibitem{exo0nu} M. Auger \textit{et al.} (EXO collaboration) Phys. Rev. Lett. \textbf{109}, 032505 (2012).
\bibitem{gerda} M. Agostini \textit{et al.} (GERDA Collaboration), Phys. Rev. Lett. \textbf{111}, 122503 (2013).

\bibitem{barabash11} A.S.\ Barabash, Phys. Atom. Nucl. \textbf{74}, 603 (2011).
\bibitem{klapdor} H.V. Klapdor-Kleingrothaus \textit{et al.}
, Phys. Lett. B \textbf{586}, 198 (2004). 


\bibitem{tello} V.\ Tello, M. Nemev\v{s}ek, F.\ Nesti, G. Senjanovi\'{c},
and F.\ Vissani, Phys. Rev.\ Lett. \textbf{106}, 151801 (2011).
\bibitem{experiment} W.-M. Yao \textit{et al.}, J. Phys. G \textbf{33}, 1 (2006).
\bibitem{rob13} R. G. H. Robertson, Mod. Phys. Lett. A \textbf{28}, 1350021 (2013).
\bibitem{del14}  S. Dell'Oro, S. Marcocci, and F. Vissani Phys. Rev. D \textbf{90}, 033005  (2014).

\bibitem{horie} H. Horie and K. Sasaki, Prog. Theor. Phys. \textbf{25},
475 (1961).
\bibitem{tomoda} T. Tomoda, Rep. Prog. Phys. \textbf{54}, 53 (1991).



 \end{thebibliography}
\end{document}